\documentclass[preprint, review, 12pt]{elsarticle}
\DeclareGraphicsExtensions{pdf,eps,jpg,png}
\usepackage{amsmath, amsfonts, amsthm, amssymb, color, soul, graphicx, subcaption, url}

\theoremstyle{definition}

\theoremstyle{remark}

\numberwithin{equation}{section}
\allowdisplaybreaks

\def\f{\frac}

\def\hf1{^\f{1}{1-\xi^2}}

\def\be{\begin{equation}}
\def\en{\end{equation}}
\def\bs{\begin{split}}
\def\es{\end{split}}
\def\ba{\begin{align}}
\def\ea{\end{align}}

\newcommand{\M}{\bf M}
\newcommand{\n }{\bf n}
\newcommand{\Q}{\bf Q}

\biboptions{sort&compress}

\usepackage{hyperref}
\begin{document}
\begin{frontmatter}
\title{Magnetic Nanoparticles in a Nematic Channel: A One-Dimensional Study}

\author[iit]{Konark Bisht}

\author[iit]{Varsha Banerjee}

\author[ub]{Paul Milewski}

\author[ub]{Apala Majumdar\corref{cor1}}
\ead{a.majumdar@bath.ac.uk}

\cortext[cor1]{Corresponding author.}

\address[iit]{Department of Physics, Indian Institute of Technology Delhi, New Delhi 110016, India.}
\address[ub]{Department of Mathematical Sciences, University of Bath, Bath BA2 7AY, UK.}

\begin{abstract}
We study a ferromagnetic suspension or a suspension of magnetic nanoparticles in an anisotropic nematic medium, in three different one-dimensional variational settings, ordered in terms of increasing complexity.  The three models are featured by a nematic energy, a magnetic energy and a magneto-nematic coupling energy and the experimentally observed patterns are modelled as local or global energy minimizers. We numerically observe polydomains with distinct states of magnetization for weak to moderate magneto-nematic coupling in our models. We demonstrate that these polydomains are stabilised by lowering the temperature (as in Mertelj et al., 2013) and that the polydomain structures lose stability as the magneto-nematic coupling increases. Some exact solutions for prototypical situations are also obtained. 
\end{abstract}
\end{frontmatter}
\section{Introduction}

Nematic liquid crystals (NLCs) are classical examples of anisotropic materials with long-range orientational ordering or ``special" directions of preferred averaged molecular alignment, referred to as directors \cite{dg}.  The presence of special directions implies that NLCs have an anisotropic response to external stimuli and are hence, highly sensitive to external fields; indeed their sensitivity to light and electric fields is one of the major reasons for their widespread applications in the display industry. However, the NLC response to magnetic fields is relatively weak e.g. fields larger than $1$ kOe are needed to reorient nematic molecules \cite{dg,VF,brochard}. In the 1970's, Brochard and de Gennes suggested that the addition of ferromagnetic particles to a nematic suspension could substantially increase the magnetic susceptibility of the suspension and there was subsequent experimental work by Rault, Cladis, and Burger \cite{brochard, radis}. There are two key factors that determine the properties of a ferromagnetic suspension - the mechanical coupling between the ferromagnetic nanoparticles and the NLC and the stability of the suspension, so that the nanoparticles do not coagulate.  In the low volume fraction limit of the suspended nanoparticles, it is reasonable to assume that the nanoparticles do not form clusters and therefore, treat the nematic order and the averaged/macroscopic magnetic moment of the suspended nanoparticles as continuous variables \cite{calderer}. 

Our work is largely motivated by the experimental results on stable ferronematic suspensions reported in \cite{Mertelj} which considered magnetic platelets suspended in an anisotropic NLC medium. In the absence of external fields, the magnetic moments of the nanoparticles interact with the NLC through surface anchoring and the magnetic moments of the nanoparticles tend to align with the nematic directors via these surface-mediated interactions, resulting in an averaged spatial magnetization in the domain \cite{calderer}. In confined geometries, the nematic director and averaged magnetization can be tailored through confinement and temperature-induced effects and our paper is based on such a model problem.

We study a suspension of ferromagnetic nanoparticles (platelets as in \cite{Mertelj}) in a NLC-filled long channel, in the dilute limit of small volume fraction. 
We take the geometry to be a two-dimensional channel
$$\Omega = \left\{(x,y) \in \mathbb{R}^2: 0\leq y \leq d; -L \leq x \leq L \right\}$$ such that $L \gg d$. We assume that the structural characteristics only vary in the $y$-direction so that the system is invariant in the $x$-direction. This is a reasonable assumption for a long, thin system with certain types of boundary conditions.
The channel dimensions are assumed to be on the micron scale and the platelets have dimensions on the nanometer scale, with platelet thickness much smaller than the diameter. 

There are two main macroscopic variables - the nematic order parameter that contains information about the orientational anisotropy and the magnetization vector, $\M$$=f M_s \mathbf{m}$, which is the spatially averaged magnetic moment of the suspended nanoparticles (where $f$ is the volume fraction, $M_s$ is the saturation value of $|\M|$ and $\mathbf{m}$ is a unit-vector). The magnetization $\M$ is induced by the alignment of the moments of the magnetic nanoparticles in an anisotropic nematic medium for a sufficiently high concentration of nanoparticles, without an external magnetic field $\mathbf{H}$. In both cases, the magnitude of the order parameter reflects the degree of ordering and the directional anisotropy is captured by the normalised order parameter. We follow the modelling approach of  Burylov and Raikher \cite{brochard1} and model the equilibrium experimentally observable configurations as local or global minimizers of an appropriately defined energy.

We can model the nematic order parameter at two levels - (i) as a two-dimensional unit-vector $\n$ $= \left(n_1, n_2 \right)$ such that $n_1^2 + n_2^2 = 1$, referred to as the Oseen-Frank director or as a (ii) two-dimensional Landau-de Gennes (LdG) $\Q$-tensor order parameter which is a symmetric, traceless $2\times 2$ matrix with two degrees of freedom \cite{luoerbanmajumdar2012}. In the Oseen-Frank model, $\n$ models the physically distinguished direction of molecular alignment in the sense that all directions perpendicular to $\n$ are physically equivalent, and in the LdG model, the $\Q$-tensor has two orthogonal eigenvectors and the eigenvector with the largest positive eigenvalue can be interpreted as the ``director". The magnitude of the LdG order parameter is a measure of the degree of orientational ordering, e.g. $\Q=0$ describes the disordered isotropic state \cite{dg}. The magnetization $\M$ is modelled as a two-dimensional vector $\M$ $= \left(M_1, M_2\right)$ with constant or variable magnitude, dependng on the modelling assumption.

We work with three different models ordered in terms of decreasing complexity - (i) the two-dimensional Landau-de Gennes model with $\M$ of variable magnitude; (ii) the Oseen-Frank model with $\M$ of variable magnitude and (iii) the Oseen-Frank model with $\M$ of constant magnitude. The first model is the most comprehensive and can account for polydomain structures in both the nematic and magnetic order; the second can only account for polydomain structures in magnetic order whilst the third cannot account for domain walls at all. We work in the absence of any magnetic fields i.e. with $\mathbf{H}=0$; this could be a good approximation to stable spatial patterns with weak magnetic fields although the asymptotics should be carefully studied. We work with Dirichlet boundary conditions for the nematic order parameter and $\M$ on the bounding surfaces i.e. treat this to be a one-dimensional problem with two bounding surfaces. Dirichlet conditions for the nematic order parameter are widely used and can usually be induced by an appropriate treatment of the boundaries. We largely choose Dirichlet conditions for the nematic order parameter so as to enforce planar boundary conditions on one surface and homeotropic/normal boundary conditions on the other surface. The nanoparticles on the boundaries are treated so as to induce a fixed alignment with respect to the nematic molecules and this fixed alignment of the long axes of the nanoparticles translates to Dirichlet conditions for $\M$ - this corresponds to the strong anchoring limit of the coupling energy proposed by Burylov and Raikher.

In all cases, the free energy has three contributions: the nematic energy, the magnetic energy and the magneto-nematic coupling energy proposed by  Burylov and Raikher. 
The coupling energy originates from the NLC coupling to the ferromagnetic nanoparticles through surface anchoring and we work in the soft anchoring limit, as proposed by  Burylov and Raikher, which can account for a rich variety of anchoring and physically relevant solutions. The precise nature of the anchoring depends on the properties of the NLC medium and the nanoparticles but we assume that the nematic molecules prefer to co-align with the long axes of the nanoparticles. 
For the most comprehensive model based on the LdG theory for NLCs, the free energy consists of a LdG energy (which is the sum of a bulk potential that enforces a preferred degree of orientational anisotropy in the bulk and a one-constant elastic energy that penalizes spatial inhomogeneities in the nematic order parameter), an analogous magnetic energy which is the sum of a potential that enforces a preferred value of $|\M|$ in the bulk and a one-constant elastic energy density and the magneto-nematic coupling energy as described above. For the second model, referred to as the Oseen-Frank model with $\M$ of variable magnitude, the nematic order parameter is a single distinguished direction that describes the preferred direction of alignment of the nematic molecules and we can recover the second model from the first in the limit of vanishing elastic constant for the LdG energy. The third model, referred to as the Oseen-Frank model with $\M$ of constant magnitude, can be recovered from the LdG model in the limits of vanishing elastic constants for both the nematic and magnetic energies. In this limit, the energy minimization can be viewed as a constrained minimization problem, where the admissible configurations are minima of the nematic and magnetic bulk potentials respectively and these bulk minimizers have constant magnitude i.e. we minimize over nematic and magnetic order parameters of constant magnitude corresponding to minimizers of the bulk potentials and hence, the nematic order parameter is the Oseen-Frank director of unit length and the magnetic order parameter, $\M$, is taken to have constant magnitude.  

The main emphasis of our work is a numerical computation of the stable configurations in this coupled system using variational methods, with special emphasis on disordered nematic regions with $\Q=0$ and regions of zero $|\M|$.
In our modelling framework, the disordered nematic regions or the regions of zero $|\M|$ define phase boundaries between regions of distinct nematic order and distinct magnetizations. Indeed, they qualitatively explain the experimentally observed polydomains of distinct magnetizations in absence of external magnetic field observed in the paper by \cite{Mertelj}. We study the stability of these polydomains as a function of the NLC elastic constant, the parameters of the magnetic energy and importantly, the strength of the magneto-nematic coupling. In fact, the stability and persistence of these polydomains can be tuned by the geometry, the boundary conditions, the NLC parameters, the properties of the nanoparticle and the mechanical coupling between the two effects and our study is only a first step on those lines. As a by-product of the equilibrium analysis in the three frameworks, we employ a simple gradient-flow model to study the dynamic evolution of the polydomains as a function of the temperature. It is expected that polydomains are stabilised at lower temperatures and the ``quenching" or rapid cooling in absence of external magnetic field can efficiently order the polydomains, as corroborated by the experimental observations in \cite{Mertelj}. We provide a qualitative understanding of these effects. The methods in this paper can be readily generalised to experiments with magnetic fields \cite{Mertelj2} to qualitatively and quantitatively model the magneto-optic response of ferronematics \cite{sluckin}. 

The paper is organised as follows. In Section \ref{energy1}-\ref{energy3}, we study equilibrium solutions and how they can be manipulated by the model parameters (with emphasis on the strength of the magneto-nematic coupling) in three different frameworks. In Section \ref{energy3}, we numerically study the stability of the polydomains and provide a numerical demonstration of the quenching effect reported in \cite{Mertelj}. We present some conclusions and directions for future work in Section \ref{conclusion}.

\section{The Oseen-Frank Model with Constant Magnetization}
\label{energy1}

As stated in the Introduction, we study the stable spatial patterns as local or global minimizers of an appropriately defined continuum energy in three different frameworks, ordered in terms of increasing complexity. The first model is the simplest one with an Oseen-Frank director and a magnetization vector of constant magnitude, with zero applied magnetic field.
The simplest form of the corresponding free energy density per unit area, for $\n = \left(\cos \varphi, \sin \varphi \right)$, $\M$ $= M \left(\cos\psi, \sin\psi \right)$ (with constant $M$) is \cite{brochard1} -
\begin{equation}
\label{eq:1}
f_1 \left(\varphi, \psi \right): = \frac{K}{2} \left(\frac{d \varphi}{ d y}\right)^2 + \frac{\kappa}{2} \left(\frac{d \psi}{ d y}\right)^2
- \frac{\gamma M^2}{2} \cos^2\left(\varphi - \psi \right) 
\end{equation}
where $K>0$ is the usual Oseen-Frank elastic constant in the one-constant approximation, $\kappa>0$ is an analogous elastic constant penalizing spatial inhomogeneities in $\M$ and $\gamma>0$ is a coupling constant that depends on subtle anchoring properties of the nematic molecules on the nanoparticle surface, the nanoparticle concentration and nanoparticle sizes \cite{brochard1}. The coupling energy is minimized for $\n$ and $\M$ parallel to one another, since $\gamma>0$; this is a system-dependent property and one could conceive of situations where the coupling energy is minimized for $\n$ and $\M$ perpendicular to one another etc. The framework is easily adapted to other choices of the magneto-nematic coupling energy.

We non-dimensionalise (\ref{eq:1}) by using the following scalings and dimensionless variables- $y^{'}= y/d$, $f_1^{'} = f_1/(\gamma M^2/2)$, $c_1 = (\gamma d^2 M^2)/(2 K)$, $c_2 = (\gamma d^2 M^2)/(2\kappa)$ so that the re-scaled energy density $f_1^{'}$ is given by
\begin{equation}
\label{eq:2}
f_1^{'} = \frac{1}{2 c_1}\left(\frac{d \varphi}{ d y^{'}}\right)^2 + \frac{1}{2 c_2}\left(\frac{d \psi}{ d y^{'}}\right)^2 - \cos^2 \left(\varphi - \psi \right).
\end{equation}
Dropping the primes, the corresponding Euler-Lagrange equations can be written as:
\begin{eqnarray}
\label{eq:3}
&& \frac{d^2 \varphi}{dy^2} = c_1 \sin 2\left(\varphi - \psi \right)\nonumber \\ && \frac{d^2 \psi}{dy^2} =- c_2 \sin 2\left(\varphi - \psi \right)
\end{eqnarray}
subject to Dirichlet boundary conditions for $\varphi$ and $\psi$ on $y=0$ and $y=1$. We can make two immediate comments on the solutions of (\ref{eq:3}); namely that for $c_1$ and $c_2$ small or equivalently for weak magneto-nematic coupling, the solutions can be well approximated by solutions of $
d^2 \varphi/dy^2 = d^2 \psi/dy^2 = 0$ with the magneto-nematic coupling term as a small perturbation. In the opposite limit of large $c_1$ and $c_2$, $\varphi \approx \psi$ almost everywhere to minimize the coupling energy in (\ref{eq:1}). We can also note that, if ($\bar{\varphi}$, $\bar{\psi}$) is the solution set of (\ref{eq:3}), then  -($\bar{\varphi}$, $\bar{\psi}$) is also a solution for the equations. According to reported values in \cite{Mertelj2}, the elastic constant $\kappa$ is usually several orders of magnitude smaller than $K$ and hence, we would expect $c_1 \ll c_2$ for physically relevant scenarios.

We illustrate these concepts below with some numerical computations. All numerical computations are carried out in COMSOL \cite{cm} and MATLAB.

\begin{figure}[!htbp]
	\centering
	\includegraphics[scale = 0.15]{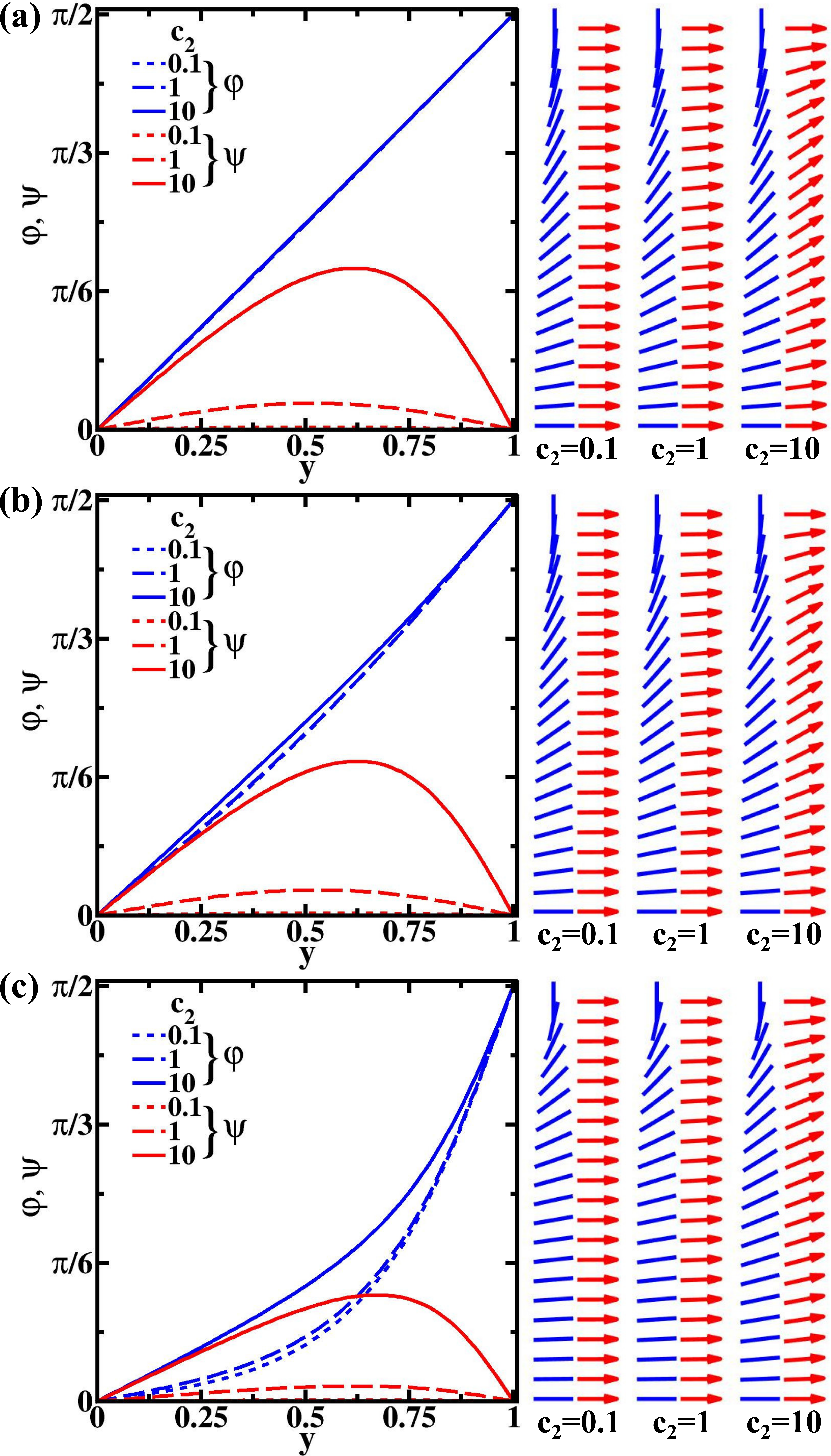}
	\caption{Solutions of (\ref{eq:3}) under boundary conditions: $\varphi = 0$ at $y=0$, $\varphi = \pi/2$ at $y=1$ and $\psi =0$ at $y=0$, $1$. The plots are for (a) $c_1 = 0.1$, (b)  $c_1 = 1$ and (c) $c_1 = 10$. For each value of $c_1$, the solutions are obtained for three different values of $c_2 = 0.1$, 1 and 10. The arrow plots show variation in $\mathbf{n}$ (blue) and $\mathbf{M}$ (red) along the $y$-axis.}
	\label{fig:1}
\end{figure}
\begin{figure}[!htbp]
	\centering
	\includegraphics[scale = 0.15]{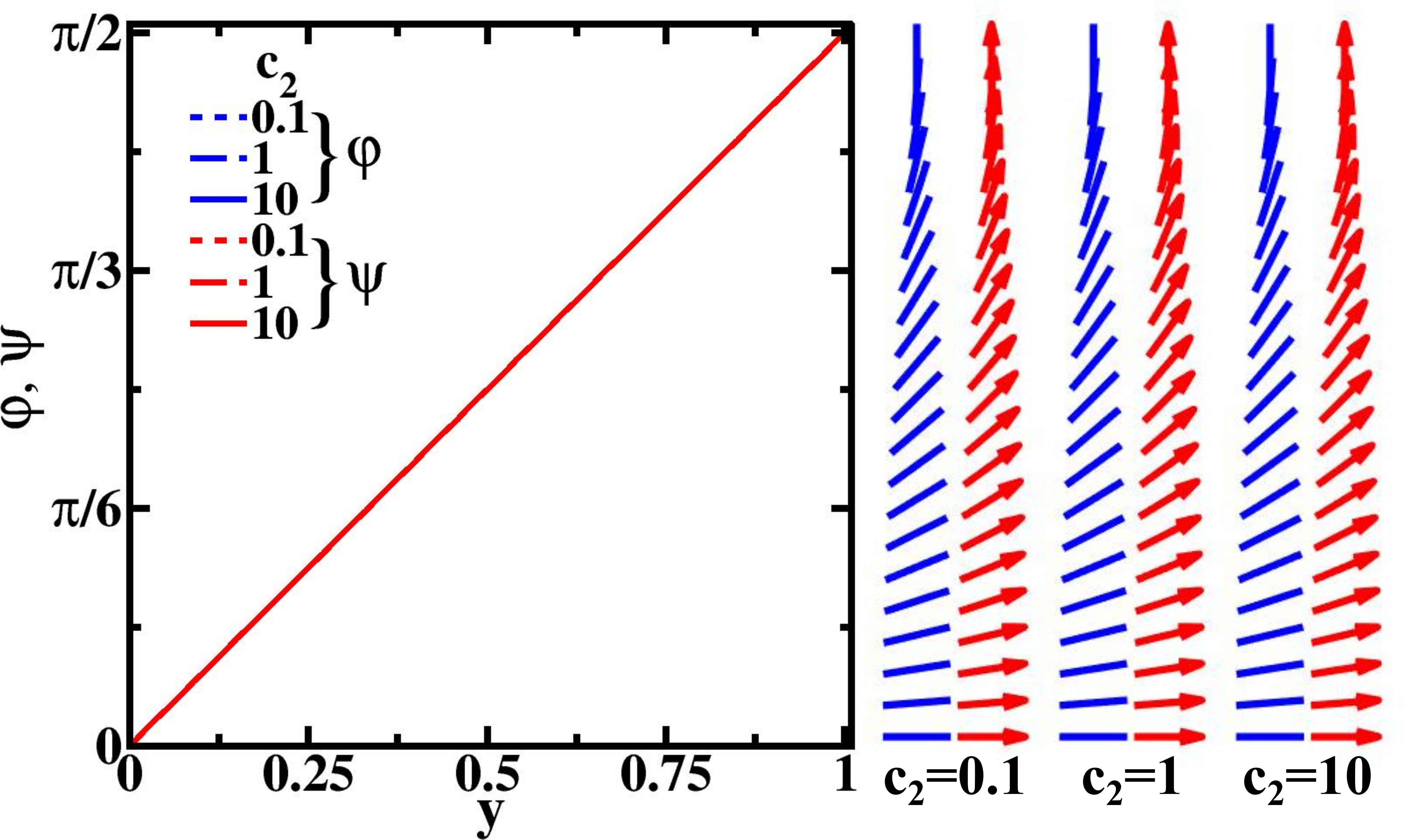}
	\caption{Solutions of (\ref{eq:3}) under boundary conditions: $\varphi = \psi = 0$ at $y=0$ and $\varphi = \psi = \pi/2$ at $y=1$. The plots are for $c_1 = 10$ and $c_2 = 0.1$, 1 and 10. We obtain linear solutions such that $\varphi= \psi$ for all values of $c_1$ and $c_2$.}
	\label{fig:2}
\end{figure}

In Figure \ref{fig:1}, we numerically solve equations (\ref{eq:3}) subject to
$\varphi = 0$ on $y=0$, $\varphi = \pi/2$ on $y=1$ and $\psi = 0$ on $y=0, 1$. This describes a situation with planar nematic anchoring,  $\n$ $= (1, 0)$ on $y=0$ and homeotropic anchoring, $\n$ $= (0,1)$ on $y=1$. The corresponding boundary conditions for $\M$ are $\M$ $ = (1,0)$ on $y=0$ and $y=1$.
We consider three different values of $c_1$ and for each value of $c_1$, we plot $\varphi$ and $\psi$ for three different values of $c_2$. For small values of $c_1$, the profile of $\varphi$ is linear and for small values of $c_2$, $\psi$ is approximately zero everywhere. For increasing values of $c_1$ and $c_2$, the profiles of $\varphi$ and $\psi$ approach each other asymptotically near $y=0$; however, they cannot agree everywhere because of the incompatible Dirichlet conditions at $y=1$. In Figure \ref{fig:2}, we plot the solutions for $\varphi$ and $\psi$ with $\varphi = \psi = 0$ at $y=0$ and $\varphi = \psi = \pi/2$ at $y=1$. It is relatively straightforward to check that we have a branch of solutions with $\varphi(y) = \psi(y)$ for $0\leq y \leq 1$, for all values of $c_1$ and $c_2$, as illustrated by the numerical simulations.

One can also derive exact solutions for the ordinary differential equations (\ref{eq:3}) as shown below. Let $\theta = \varphi- \psi$ and $q = c_2 \varphi + c_1 \psi$; then
\begin{eqnarray}
\label{eq:4}
&& \frac{d^2 \theta}{dy^2} = \left( c_1 + c_2 \right) \sin 2 \theta \nonumber \\ && \frac{d^2 q}{dy^2} =0
\end{eqnarray}
subject to Dirichlet conditions, $\theta (0) = \varphi(0) - \psi(0)$ and $\theta(1) = \varphi(1) - \psi(1)$ and $q(0) = c_2 \varphi(0) + c_1 \psi(0)$, $q(1) = c_2\varphi(1) + c_1 \psi(1)$.
It is easily checked that
$$ q = ay +b $$
for
$$ a =  \left(q(1) - q(0) \right); \quad b = q(0). $$
Regarding $\theta$, one can check that the first ordinary differential equation is equivalent to
\begin{equation}
	\label{eq:5}
	\left( \frac{d \theta}{dy} \right)^2 + 2\left(c_1 + c_2 \right) \cos^2\theta = \mathcal{C}
\end{equation}
for some constant $\mathcal{C}$ determined by
\begin{equation}
	\label{eq:6}
	\int_{\theta(0)}^{\theta(1)} \frac{d \theta}{\sqrt{ \mathcal{C} - 2(c_1 + c_2)\cos^2\theta }} = \pm 1,
\end{equation}
the sign depending on whether we seek monotonically increasing or decreasing solutions for $\theta$ (i.e. if $\theta(0) < \theta(1)$ or if $\theta(0)> \theta(1)$). If $\theta(0) = \theta(1)$, then there may be an intermediate stationary point and that case can be dealt with by similar methods by assuming that the stationary point is located at $y=1/2$ and we have a symmetric profile.
Once $\mathcal{C}$ is fixed by (\ref{eq:6}), we can compute $\theta$ in terms of special functions such as complete elliptic integrals $\mathcal{K}(m)$ and Jacobi elliptic function sn(Y; m) \cite{elliptic,AB}. Defining $m = 2(c_1+c_2)/\mathcal{C}$ and $Y = \mathcal{K}(m) - \sqrt{\mathcal{C}}y$, the  exact solution of (\ref{eq:5}) can be written as
\begin{equation}
	\label{eq:7}
	\theta\left(y \right) =  \arccos\left(\textrm{sn}\left(Y; m\right)\right).
\end{equation}
One can check that the exact solutions in (\ref{eq:7}) coincide with the numerical solutions computed above for the same sets of boundary conditions.
We can use similar methods to compute exact solutions for $\varphi$ and $\psi$ with multiple turning points although such solutions are expected to have higher energies than the numerically reported solutions here.

\section{The Oseen-Frank Model with Variable Magnetization}
\label{energy2}

For the second model, we employ the nematic director, $\n = \left(\cos\varphi, \sin\varphi \right)$ and the magnetization vector $\M$ $ = \left(M_1, M_2 \right)$ without the assumption of $|\M|$  $=M$ constant. One potential benefit of this approach is that it allows us to study nodal lines of $\M$ or regions of zero magnetization and novel solutions for $\varphi$.
The corresponding free energy density is
\begin{eqnarray}
\label{eq:8}
&f_2\left(\varphi, M_1, M_2 \right):=&\frac{K}{2} \left(\frac{d \varphi}{ d y}\right)^2+ \frac{\kappa}{2}\left(\left(\frac{d M_1}{ d y}\right)^2 + \left(\frac{d M_2}{ d y}\right)^2 \right) + \frac{\alpha}{2}{\M \cdot \M} +\nonumber \\&&+ \frac{\beta}{4} \left(\M\cdot \M\right)^2-\frac{\gamma}{2}
\left(M_1 \cos\varphi + M_2 \sin\varphi \right)^2.
\end{eqnarray}

Here $K>0$ is the nematic elastic constant, $\alpha$ is a temperature-dependent parameter and $\beta>0$ is  material dependent constant that dictate the preferred value of $|\M|$ in the bulk (see \cite{landau} for similar examples of Landau energies), $\kappa>0$ is an elastic constant that penalises spatial inhomogeneities in $\M$ and $\gamma>0$ is the magneto-nematic coupling constant as before. We work with $\alpha<0$ so that a non-zero $|\M|$ is preferred in the bulk. We re-scale the energy density using $y^{'} =y/d$, $M_1^{'} = c_m M_1$, $M_2^{'} = c_m M_2$ where $c_m = \sqrt{\beta/|\alpha|}$ to get the re-scaled energy density (omitting the hyphens for brevity)
\begin{eqnarray}
\label{eq:9}
&f_2 \left(\varphi, M_1, M_2 \right):=&\dfrac{\ell_1}{2}\left(\dfrac{d \varphi}{ d y}\right)^2 + \dfrac{1}{c}\left[\ell_2 \left(\left(\dfrac{d M_1}{ d y}\right)^2 + \left(\dfrac{d M_2}{ d y}\right)^2 \right)- {\M\cdot\M} +\right. \nonumber\\&&\left. +\frac{1}{2}\left(\M\cdot\M\right)^2 \right]- \left(M_1 \cos\varphi + M_2\sin\varphi\right)^2
\end{eqnarray}
where $$ c= \frac{\gamma}{|\alpha|}; \quad \ell_1 = \frac{2\beta K}{|\alpha|\gamma d^2}; \quad \ell_2 = \frac{\kappa}{|\alpha|d^2}. $$
The corresponding Euler-Lagrange equations are: 
\begin{eqnarray}
\label{eq:10}
&& \ell_1 \frac{d^2 \varphi}{dy^2} =  \left(M_1^2 - M_2^2 \right)\sin 2 \varphi - 2 M_1 M_2 \cos 2\varphi,\nonumber \\ && \ell_2 \frac{d^2 M_1}{dy^2} = - M_1 + \left(M_1^2 + M_2^2 \right)M_1 - c \cos\varphi \left(M_1 \cos\varphi + M_2 \sin\varphi \right),\nonumber \\ && \ell_2 \frac{d^2 M_2}{dy^2} = - M_2 + \left(M_1^2 + M_2^2 \right)M_2 - c \sin\varphi \left(M_1 \cos\varphi + M_2 \sin\varphi \right).
\end{eqnarray}
We can derive an alternative formulation by using $M_1 = M \cos\psi$, $M_2 = M\sin\psi$, the Euler-Lagrange equations (\ref{eq:10}) can be re-written as
\begin{eqnarray}
\label{eq:11}
&& \ell_1 \frac{d^2 \varphi}{dy^2} = M^2 \sin 2\left(\varphi - \psi \right), \nonumber \\ && \ell_2 \left(M \frac{d^2 \psi}{dy^2} + 2 \frac{dM}{dy}\frac{d\psi}{dy} \right) =- \frac{c M}{2} \sin 2 \left(\varphi - \psi \right),\nonumber \\ && \ell_2 \left(\frac{d^2 M}{dy^2} -  M \left(\frac{d\psi}{dy}\right)^2 \right) =- M + M^3 - \frac{cM}{2}\left(1 + \cos 2\left(\varphi - \psi\right)\right).
\end{eqnarray} The Ginzburg-Landau parallelism can be seen more clearly by writing $\mathbf{M} = M \exp\left( i \psi \right)$, so that the partial differential equations for $M_1$ and $M_2$ can be combined to give 
\begin{equation}
\label{eq:11b}
\ell_2 \mathbf{M} _{yy} = -\left(1 + \frac{c}{2} \right)\mathbf{M}  + |\mathbf{M} |^2 
\mathbf{M}  - \frac{c}{2} \mathbf{M}  \exp\left(2 i \left(\phi - \psi \right)\right).
\end{equation}

One can check that for Dirichlet conditions, $\varphi(0) = \psi(0)$, $\varphi(1) = \psi(1)$ and $M(0) = M(1) = M^*$, we have a branch of solutions of (\ref{eq:11}) given by
\begin{eqnarray}
\label{eq:12}
\varphi = \psi = a y + b;\quad M = M^* \textrm{ constant}
\end{eqnarray}
with constants $a$ and $b$ being determined by the boundary conditions and $M^*$ being determined by the roots of the cubic polynomial
\begin{equation}
\label{eq:13}
M\left( 1 - a^2 \ell_2 \right) - M^3 + c M = 0
\end{equation}
such that
$$ M^* = 0, \pm \sqrt{1 + c - \ell_2 a^2 }$$ provided $c> \ell_2 a^2 - 1$. 

\begin{figure}[!htbp]
	\centering
	\includegraphics[scale = 0.15]{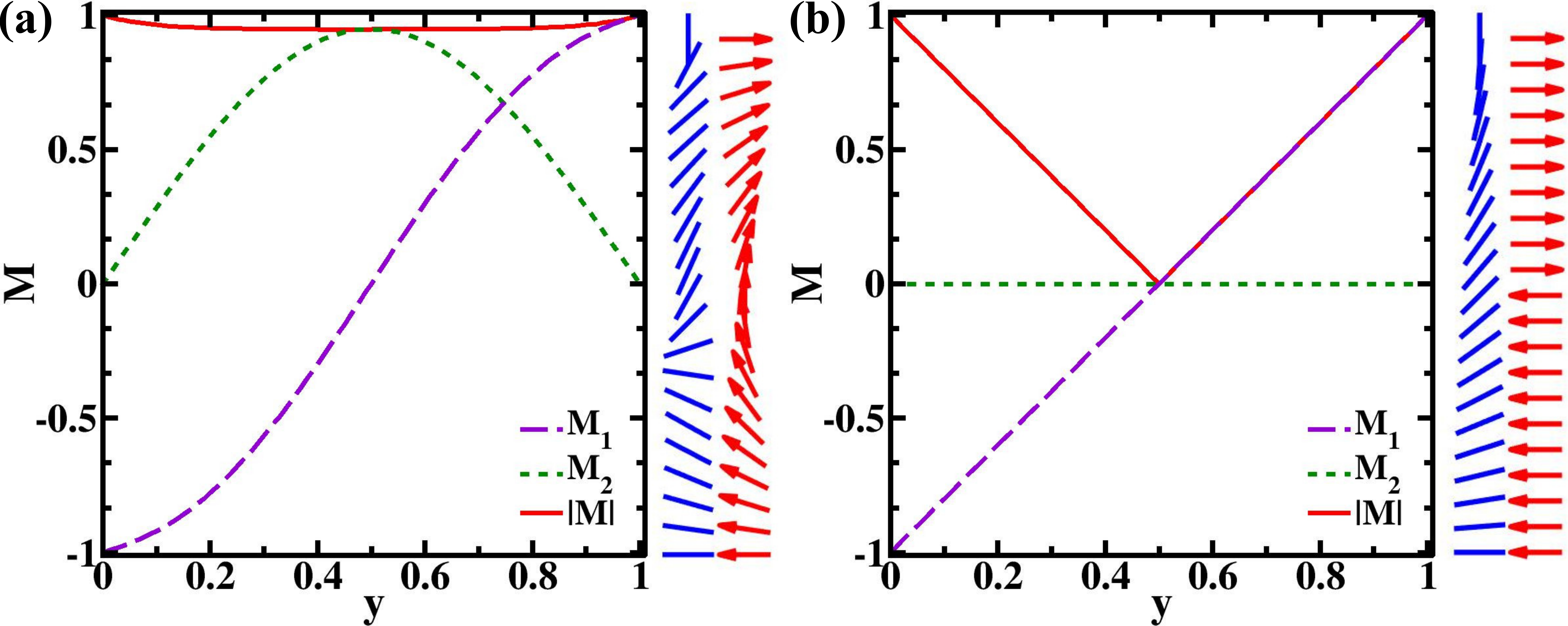}
	\caption{Solutions of (\ref{eq:10}) under boundary conditions: $\varphi(0) = 0$, $\varphi(1) = \pi/2$, $\M$$(0)= (-1, 0)$ and $\M$$(1)= (1,0)$ for two distinguishing limits (a) Ginzburg-Landau limit ($\ell_1 = \ell_2 = 0.01$ and $c=0.001$) and (b) Laplace limit  ($\ell_1 = \ell_2 = 10$ and $c=10$). The arrow plots show variation in $\mathbf{n}$ (blue) and $\mathbf{M}$ (red) along the $y$-axis.}
	\label{fig:ginzburg}
\end{figure}
There are two distinguished limits to be considered. The first limit is the Ginzburg-Landau limit for when $\ell_1, \ell_2 \ll 1$ and $c$ is very small. This is analogous to the $\epsilon \to 0$ limit in the Ginzburg-Landau theory for superconductivity \cite{brezis_bethuel_helein} that coerces $|\M| \to 1$ almost everywhere and $M_1$ and $M_2$ to be solutions of
\begin{equation}
\label{eq:14}
\frac{d^2 M_1}{dy^2}= \frac{d^2 M_2}{dy^2} = 0
\end{equation}
subject to the boundary conditions and the unit length constraint. The values for $\alpha$ and $\beta$ are not frequently reported but $\gamma$ can be as large as $100$ in appropriate units (see \cite{Mertelj2}) whilst $d \sim 10^{-6} m$, $K$ and $\kappa$ have very small magnitudes in their respective units, so that we expect $\ell_1, \ell_2 \ll 1$ and the Ginzburg-Landau limit to be the physically relevant limit. 
In Figure \ref{fig:ginzburg}(a), we plot $\M$ for $\varphi(0) = 0$, $\varphi(1) = \pi/2$, $\M$$(0)= (-1, 0)$ and $\M$$(1)= (1,0)$ for $\ell_1 = \ell_2 = 0.01$ and $c=0.001$. We clearly see that $|{\M}| \approx 1$ almost everywhere, the profile of $M_1$ is approximately linear and $M_2$ has a parabolic profile to compensate for the fact that $M_1$ cannot have unit norm everywhere. The profile of $\varphi$ is not linear since the dominant term is the forcing term $(M_1^2 - M_2^2)\sin 2\varphi + 2M_1 M_2 \cos 2\varphi$, which is non-zero.

\begin{figure}[!htbp]
	\centering
	\includegraphics[scale = 0.118]{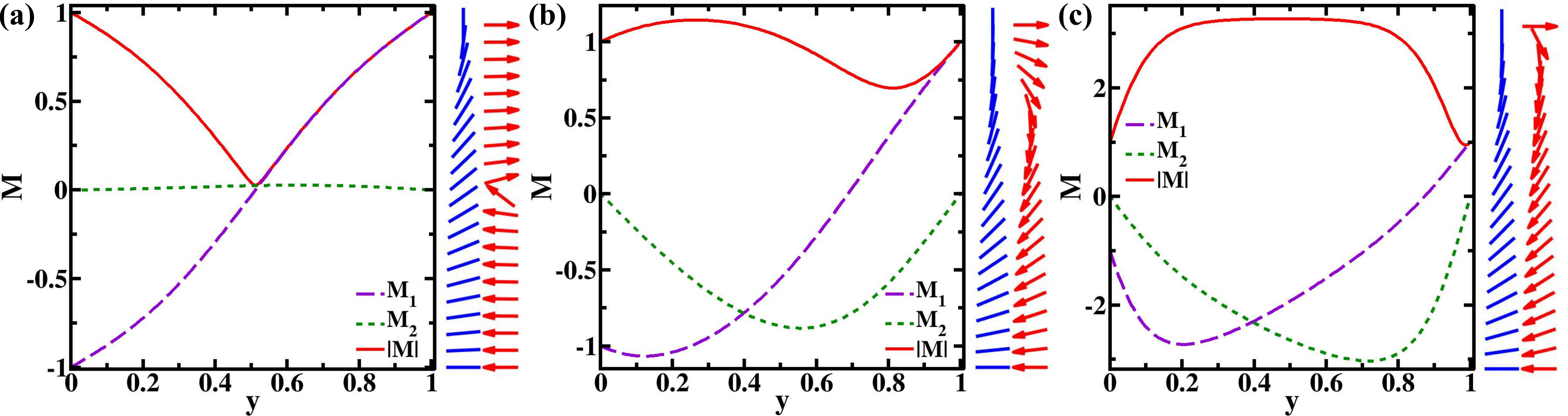}
	\caption{Solutions of (\ref{eq:10}) under boundary conditions $\varphi(0) = 0, \varphi(1) = \pi/2$, $\M$$(0) = (-1, 0)$ and $\M$$(1) = (1,0)$.  The solutions are obtained for $\ell_1 = \ell_2 = 0.1$ and three different values of $c$: (a) $c =0.1$, (b) $c =1$ and (c) $c =10$.}
	\label{fig:model2_1}
\end{figure}
The second distinguished limit is the Laplace limit for large $\ell_1$ and $\ell_2$. In this limit, the leading order solutions are solutions of the second order ordinary differential equations; $$ \frac{d^2 \varphi}{dy^2} =   \frac{d^2 M_1}{dy^2}= \frac{d^2 M_2}{dy^2} = 0$$
subject to the imposed boundary conditions. The numerics suggest that the Laplace limit is valid for $\ell_1, \ell_2 \geq \max\left\{c_1, c_2, 1\right\}$ although this needs to be more systematically studied. In Figure \ref{fig:ginzburg}(b), we plot the solutions for $\M$ with $\ell_1 = \ell_2 = 10$ and $c=10$, with $\varphi(0) = 0$, $\varphi(1) = \pi/2$, $\M$$(0) = (-1, 0)$ and $\M$$(1) = (1,0)$. We see that the solution profile for $\varphi$ is linear as expected under the Laplace limit, as is the plot for $M_1$ whereas $M_2$ is identically zero within numerical approximation. In this limit, there is a distinct wall of zero magnetization at $y \simeq 1/2$ which cannot be captured by the Oseen-Frank model with constant magnetization. 

In Figure \ref{fig:model2_1}, we plot the $\M$ profiles for $\ell_1=\ell_2 = 0.1$ and three different values of $c$, for $\varphi(0) = 0$, $\varphi(1) = \pi/2$, $\M$$(0) = (-1, 0)$ and $\M$$(1) = (1,0)$. We see the signature of the Laplace limit for $|\ell_1| =|\ell_2| = |c|$ and as $c$ increases relative to $\ell_1, \ell_2$, $M_2$ assumes non-trivial profiles and the $\M$ profiles get increasingly distorted.

\begin{figure}[!htbp]
	\centering
	\includegraphics[scale = 0.15]{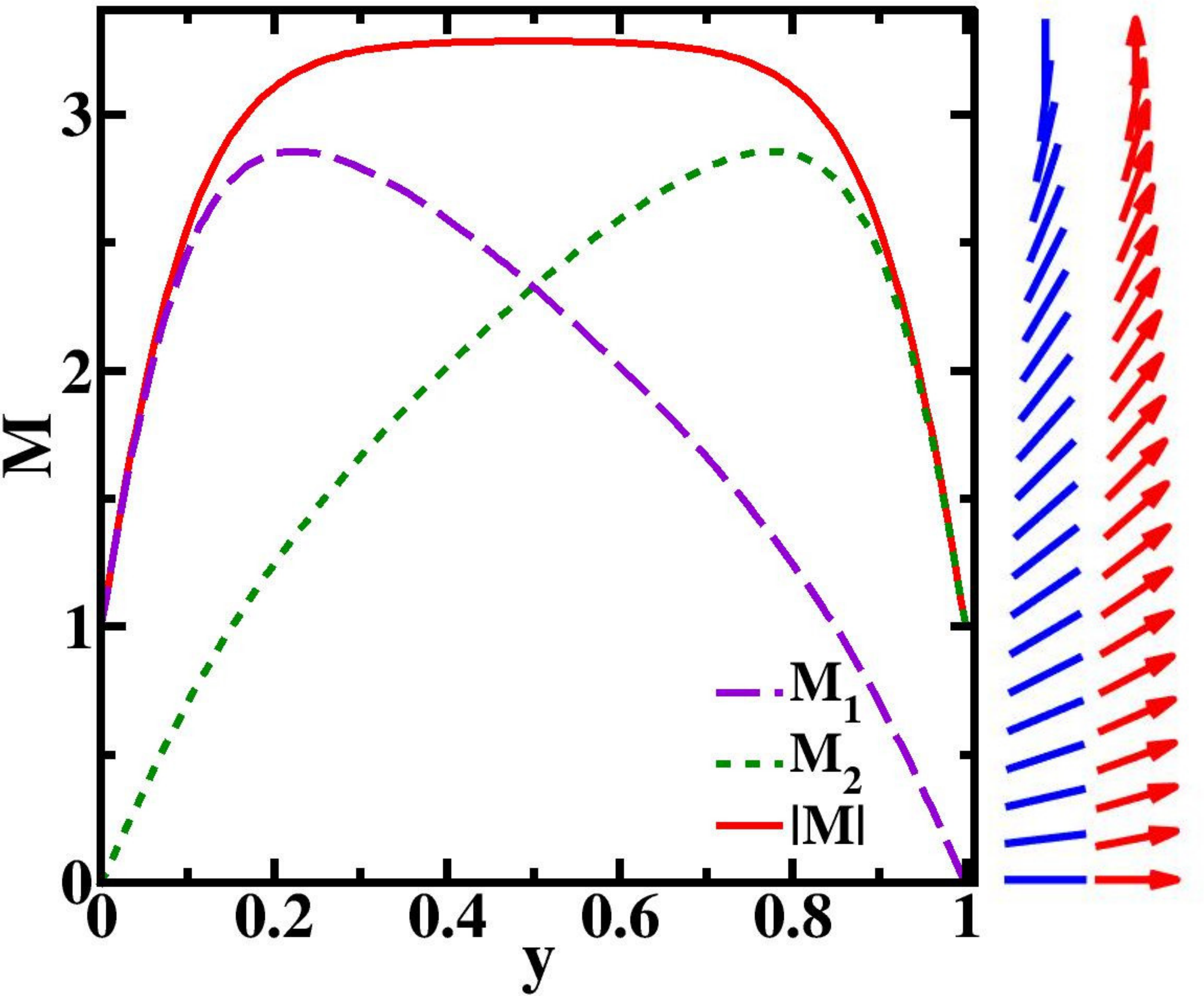}
	\caption{Solutions of (\ref{eq:10}) under boundary conditions $\varphi(0) = 0, \varphi(1) = \pi/2$, ${\M}$$(0) = (1, 0)$ and ${\M}$$(1) = (0,1)$ for $\ell_1 = \ell_2 = 0.1$ and $c=10$. }
	\label{fig:model2_3}
\end{figure}
In Figure \ref{fig:model2_3}, we plot the $\varphi$ and $\M$ profiles for $\varphi(0) = 0, \varphi(1) = \pi/2$; $\M$$(0) = (1, 0)$ and $\M$$(1) = (0,1)$. This corresponds to the same boundary conditions for $\varphi$ and $\psi$ at $y=0$ and $y=1$. We observe that  $|\M|$ tends to a constant in the middle of the sample although we have not carried out exhaustive simulations to deduce if this may be a generic feature of the solutions. In general, there is no reason why $|\mathbf{M}|$ should be a constant or why $|\M|$$ = M^*$ (where $M^*$ is given by (\ref{eq:13})) on the boundaries.  A further possibility is that solutions with constant $|\M|$ may not be energetically minimal and the numerical algorithms can converge to solutions with lower energies. We do not explore the class of solutions defined by (\ref{eq:12}) and (\ref{eq:13}) in great detail in this manuscript. 

\section{The Landau-de Gennes Model with Variable Magnetization}
\label{energy3}

In the Landau-de Gennes framework, we model the state of the nematic by a tensor order parameter which contains information about both the degree and directions of orientational ordering. We work in a reduced two-dimensional Landau-de Gennes framework for which the order parameter is a symmetric, traceless $2\times 2$ matrix \cite{luoerbanmajumdar2012} i.e.
\begin{equation}
\label{eq:15}
{\Q} = s\left(\mathbf{n} \otimes \mathbf{n} - \frac{I_2}{2} \right)
\end{equation}
where $\n = \left(\cos\varphi, \sin\varphi \right)$, $s$ is a scalar order parameter that measures the degree of order about $\n$ and $I_2$ is the $2\times 2$ identity matrix. This order parameter has two degrees of freedom
\begin{eqnarray}
\label{eq:16}
&& Q_{11} = \frac{s}{2}\cos 2 \varphi;\quad Q_{12} = \frac{s}{2}\sin 2 \varphi.
\end{eqnarray}

The Landau-de Gennes free energy density with a magneto-nematic coupling term is given by
\begin{eqnarray}
\label{eq:17}
f_3\left( Q_{11}, Q_{12}, M_1, M_2 \right): = f_{nem} + \frac{\alpha}{2}|\mathbf{M}|^2 + \frac{\beta}{4}|\mathbf{M}|^4  + \frac{\kappa}{2}\left(\frac{d \mathbf{M}}{dy}\right)^2- \frac{\gamma}{2} M_i Q_{ij} M_j
\end{eqnarray}
where $Q_{ij}$ are the Landau-de Gennes order parameter components with $i,j = 1,2$. The contribution of higher order magneto-nematic coupling terms are not considered as the cubic coupling ($\sim \gamma$) is sufficient to induce magneto-nematic ordering  \cite{HP}.  We take $\alpha<0$ so that we have preferred non-zero $|\M|$ in the bulk. The nematic energy density, $f_{nem}$, is the usual Landau-de Gennes free energy density in two dimensions
\begin{eqnarray}
\label{eq:18}
f_{nem} = \frac{A}{2} |\mathbf{Q}|^2 + \frac{C}{4}|\mathbf{Q}|^4 + \frac{L}{2}\left( \frac{d\mathbf{Q}}{dy}\right)^2;
\end{eqnarray}
where $|{\Q}| = $(tr ${\Q}^2)^{\frac{1}{2}}$. We take $A<0$ so that we work with low temperatures below the supercooling temperature and $C, L$ are positive material dependent constants \cite{luoerbanmajumdar2012}. We note that the cubic term in the Landau-de Gennes bulk potential necessarily vanishes for two-dimensional $\Q$-tensors as in (\ref{eq:15}) and such reduced descriptions work well for severely confined systems, when the third dimension is much smaller than the lateral dimensions \cite{dg, luoerbanmajumdar2012}.

We employ the scalings ${\Q}^{'} = \sqrt{2C/|A|}$ ${\Q}$, ${\M}^{'} = \sqrt{\beta/|\alpha|}$ ${\M}$ and $y^{'} = y/d $ and the re-scaled energy density (dropping the tildes) is (also see \cite{luoerbanmajumdar2012})
\begin{eqnarray}
\label{eq:19}
&f_3\left( Q_{11}, Q_{12}, M_1, M_2 \right):=& \frac{1}{c_1}\left[ - \left(Q_{11}^2 + Q_{12}^2 \right) + \frac{1}{2}\left(Q_{11}^2 + Q_{12}^2 \right)^2 + \ell_1 \left(\frac{d \Q}{dy}\right)^2 \right] + \nonumber \\
&& + \frac{1}{c_2}\left[ -\left(M_1^2 + M_2^2\right) + \frac{1}{2}\left(M_1^2 + M_2 \right)^2 + \ell_2  \left(\frac{d \M}{dy}\right)^2 \right] - \nonumber \\&& - \left\{ Q_{11}\left( M_1^2 - M_2^2 \right) + 2 Q_{12} M_1 M_2 \right\}.
\end{eqnarray}
The corresponding Euler-Lagrange equations are:
\begin{eqnarray}
\label{eq:20}
&& \ell_1 \frac{d^2 Q_{11}}{dy^2} = - Q_{11} + \left(Q_{11}^2 + Q_{12}^2\right) Q_{11} - \frac{c_1}{2}\left(M_1^2 - M_2^2 \right) \nonumber \\
&& \ell_1 \frac{d^2 Q_{12}}{dy^2} = - Q_{12} + \left(Q_{11}^2 + Q_{12}^2\right) Q_{12} - c_1 M_1 M_2 \nonumber \\
&& \ell_2 \frac{d^2 M_1}{dy^2} = - M_1 + \left(M_1^2 + M_2^2\right) M_1 - c_2 \left( Q_{11} M_1 + Q_{12} M_2 \right) \nonumber \\
&& \ell_2 \frac{d^2 M_2}{dy^2} = - M_2 + \left(M_1^2 + M_2^2\right) M_2 - c_2 \left( Q_{12} M_1 - Q_{11} M_2 \right)
\end{eqnarray}
where
$$ \ell_1 = \frac{L}{d^2 |A|}; \ell_2 = \frac{\kappa}{d^2 |\alpha|}; c_1 = \frac{\gamma |\alpha|}{\beta |A|} \sqrt{\frac{C}{2|A|}}; c_2 = \frac{\gamma}{|\alpha|} \sqrt{\frac{|A|}{2C}}. $$

By making the transformations: $Q_{11} = |{\Q}|\cos\phi/\sqrt{2};  Q_{12} = |{\Q}|\sin\phi/\sqrt{2};M_{1} = M\cos\psi$; $ M_{2} = M\sin\psi$ in (\ref{eq:20}), one can verify that there exist a branch of solutions given by 
\begin{eqnarray}
\label{eq:21}
\varphi = \psi = a y + b ;\quad |{\Q}| = Q^* \textrm{ constant};\quad M = M^* \textrm{ constant}
\end{eqnarray}
for the Dirichlet conditions, $|{\Q}(0)|=|{\Q}(1)|= Q^*$, $M(0)=M(1)= M^*$ and $\phi= \psi$ at $y=0, 1$. The constants $a$ and $b$ can be set by the boundary conditions. The $Q^*$ is determined as the positive real root of cubic polynomial
\begin{eqnarray}
	\label{eq:22}
	Q^3 -pQ-\sqrt{2}q = 0,
\end{eqnarray}
 and is given by 
\begin{eqnarray}
	\label{eq:23}
	Q^* =\dfrac{2^{1/6}p}{3^{1/3}\Delta}+\dfrac{\Delta}{2^{1/6}3^{2/3}}
\end{eqnarray}
provided $27q^2>2p^3$ where $$p = 2+c_1c_2-8a^2\ell_1;\quad q = c_1(1-a^2\ell_2);\quad \Delta = \left(9q+\sqrt{3\left(27q^2-2p^3\right)}\right)^{1/3}.$$
The $M^*$ can then be obtained from the relation
\begin{eqnarray}
	\label{eq:24}
	M^* = \sqrt{\frac{c_2Q^*}{\sqrt{2}}+\frac{q}{c_1}}.
\end{eqnarray}

\begin{figure}[!htbp]
	\centering
	\includegraphics[scale = 0.15]{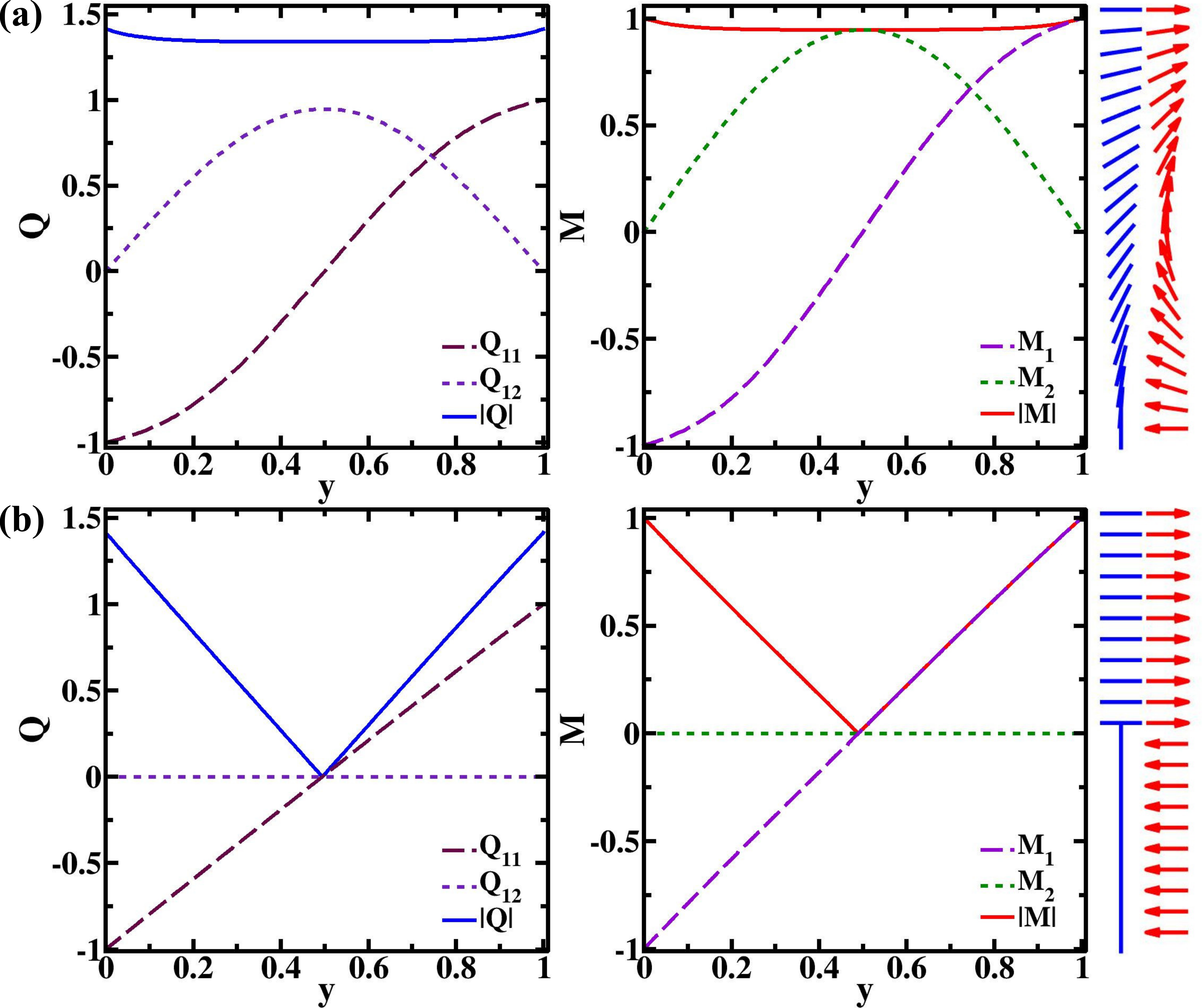}
	\caption{Solutions of (\ref{eq:20}) under boundary conditions: $Q_{11}(0) = -1$, $Q_{11}(1) = 1$, $Q_{12}(0) = Q_{12}(1) = 0$, $\M$$(0)= (-1, 0)$ and $\M$$(1)= (1,0)$ for (a) Ginzburg-Landau limit ($\ell_1 = \ell_2 = 0.01$ and $c=0.0001$) and (b) Laplace limit  ($\ell_1 = \ell_2 = 10$ and $c=10$). The arrow plots show variation in $\mathbf{n}$ (blue) and $\mathbf{M}$ (red) along the $y$-axis.}
	\label{fig:model3_1}
\end{figure}
As in Section \ref{energy2}, there are two distinguished limits for the solutions of the system (\ref{eq:20}): the Ginzburg-Landau limit with very small $\ell_1, \ell_2, c_1, c_2$ that coerces $|\mathbf{Q}|, |\mathbf{M}| \approx 1$ almost everywhere; we illustrate this in Figure \ref{fig:model3_1}(a) where we plot $\mathbf{Q} = \left(Q_{11}, Q_{12} \right)$, $\mathbf{M} = \left(M_1, M_2 \right)$ with $Q_{11}(0) = -1$, $Q_{11}(1) = 1$, $Q_{12}(0) = Q_{12}(1) = 0$, $\M$$(0)= (-1, 0)$ and $\M$$(1)= (1,0)$, for $\ell_1 = \ell_2 = 0.01$ and $c_1 = c_2= 0.0001$. 
The second distinguished limit is the Laplace limit with large $\ell_1$ and $\ell_2$
for which the solutions for $\Q$ and $\M$ are well approximated by solutions of the second order differential equation, $d^2 {\Q}/dy^2 = d^2 {\M}/dy^2 = 0$ i.e. linear profiles subject to the imposed boundary condition. In Figure \ref{fig:model3_1}(b), we plot the corresponding solution profiles for $\ell_1 = \ell_2 = 10$ and provided the re-scaled elastic constants are sufficiently large (in some cases, we only need $\ell_1,\ell_2 \geq c_1,c_2$), we see linear solution profiles in the so-called ``Laplace limit" with $Q_{12}$ and $M_2$ identically zero.
It is also relatively straightforward to check that $Q_{12} = M_2=0$ are solutions of the system (\ref{eq:20}) for all values of $\ell_1, \ell_2, c_1, c_2$, provided they are compatible with the boundary conditions.  These solutions describe domain walls that separate two regions of distinct nematic ordering or distinct magnetizations. In fact, such phase-separated solutions may offer new physical and applications-oriented perspectives. Further, for solutions with $Q_{12} = M_2 = 0$, we have a branch of solutions with $Q_{11} = M_1$ for $c_2 = c_1/2$, provided $Q_{11}$ and $M_1$ have the same boundary conditions. As in the previous section, we expect the Ginzburg-Landau limit to be the physically relevant limit based on the relative magnitudes of the physical elastic constants and the Laplace limit to be relevant for severely confined systems where $d^2$ is comparable to material correlation lengths, such as $L/|A|$ or $\kappa/|\alpha|$.

\begin{figure}[!htbp]
	\centering
	\includegraphics[scale = 0.15]{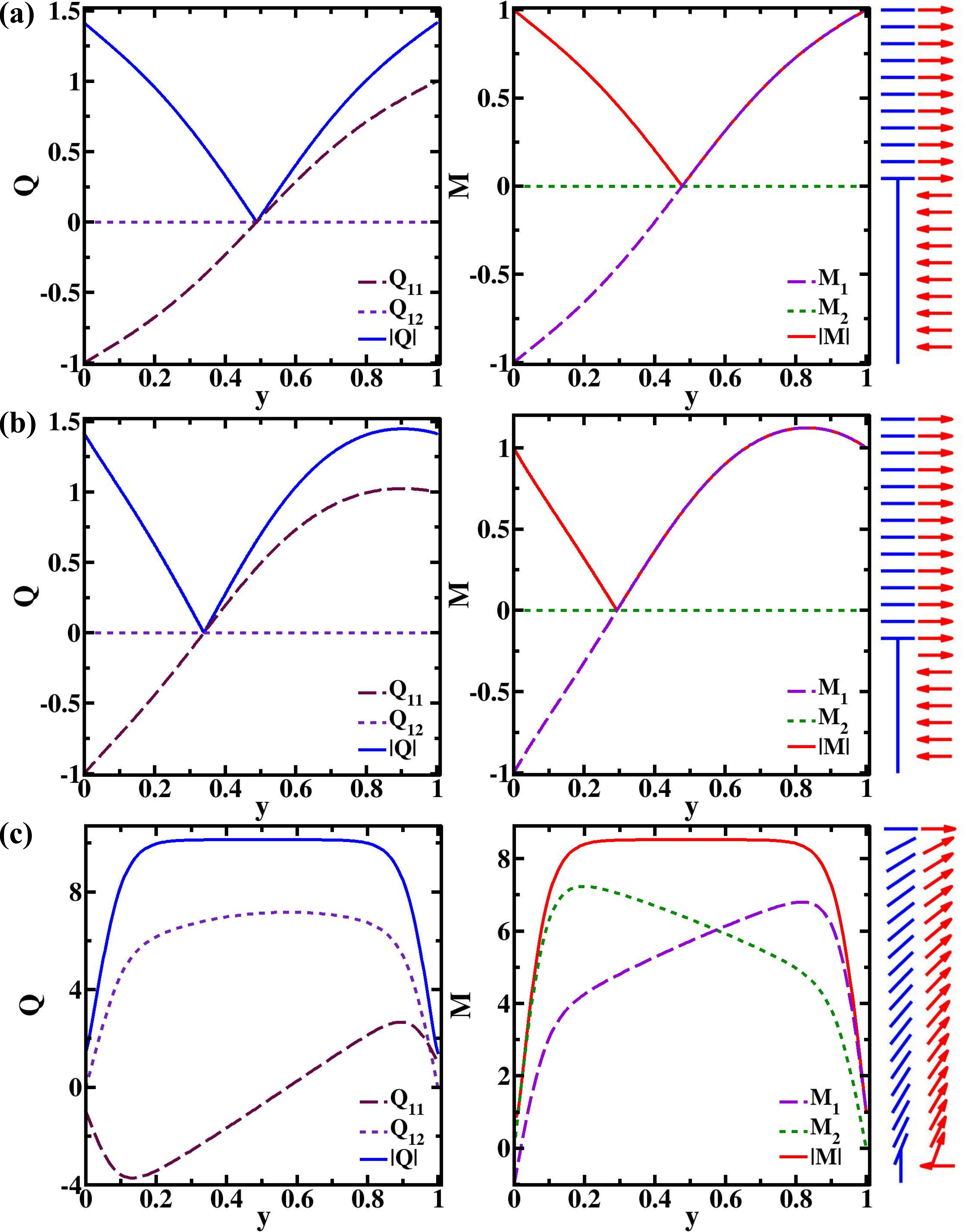}
	\caption{Solutions of (\ref{eq:20}) under boundary conditions: $Q_{11}(0) = -1$, $Q_{11}(1) = 1$, $Q_{12}(0) = Q_{12}(1) = 0$, $\M$$(0)= (-1, 0)$ and $\M$$(1)= (1,0)$. The plots are for (a) $c_1 = c_2 = 0.1$, (b)  $c_1 = c_2 = 1$ and (c) $c_1 = c_2 = 10$.}
	\label{fig:model3_2a}
\end{figure}
\begin{figure}[!htbp]
	\centering
	\includegraphics[scale = 0.15]{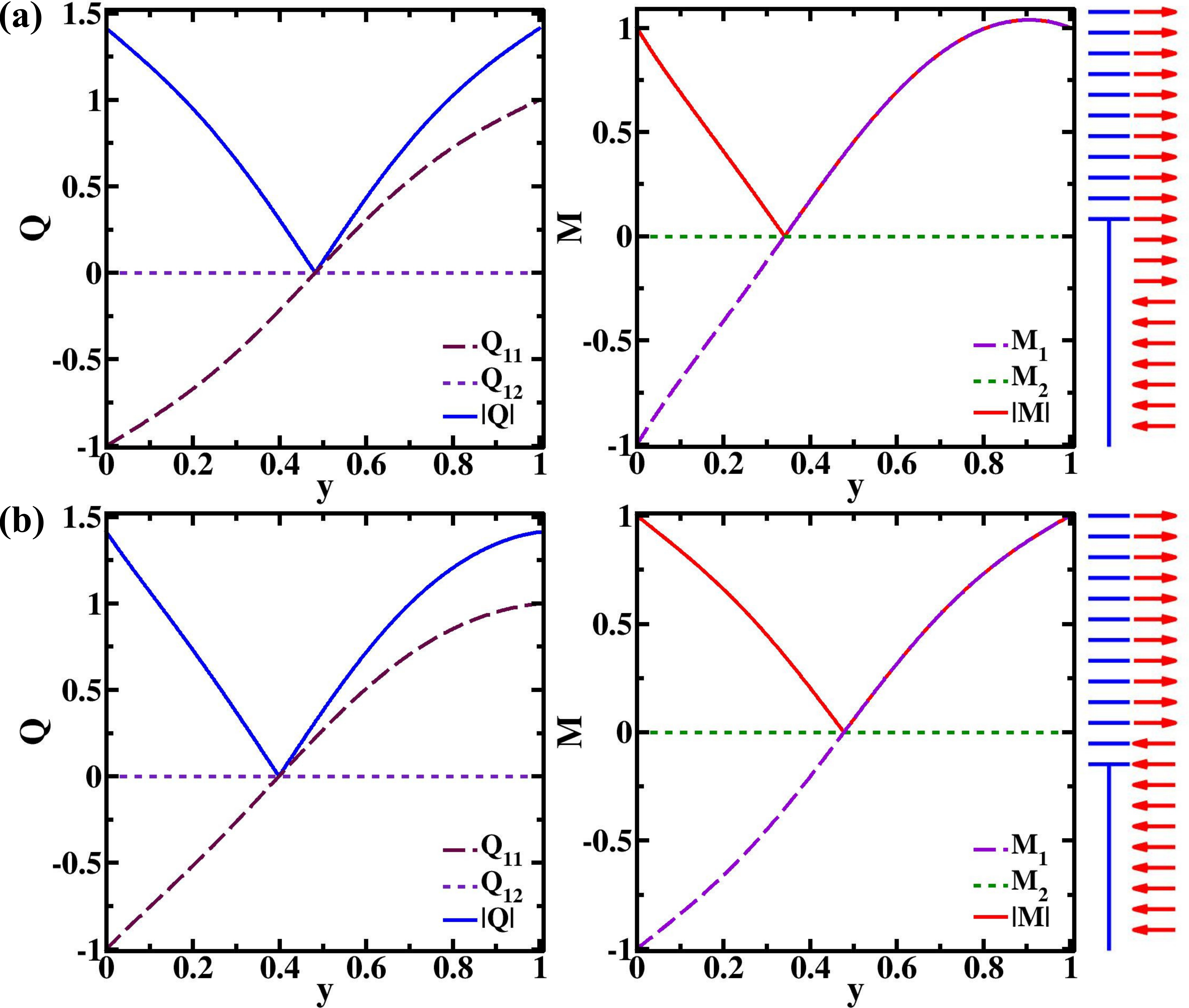}
	\caption{Solutions of (\ref{eq:20}) under boundary conditions: $Q_{11}(0) = -1$, $Q_{11}(1) = 1$, $Q_{12}(0) = Q_{12}(1) = 0$, $\M$$(0)= (-1, 0)$ and $\M$$(1)= (1,0)$. The solutions are obtained for the case $c_1\neq c_2$: (a) $c_1 = 0.1, c_2 = 1$ and (b)  $c_1 = 1$,$c_2=0.1$.}
	\label{fig:model3_2b}
\end{figure}

We illustrate these concepts in Figure \ref{fig:model3_2a} where we plot the solutions for $\Q$ and $\M$ with $c_1=c_2$ and $\ell_1 = \ell_2 =0.1$ (with $Q_{11} = -1, M_1 = -1$ at $y=0$; $Q_{11}=1, M_1 = 1$ at $y=1$, $Q_{12} = M_2 = 0$ at $y=0$ and $y=1$). We plot the solutions for the case $c_1\neq c_2$ in Figure \ref{fig:model3_2b}. We get approximately linear profiles for cases Figure \ref{fig:model3_2a}(a)-(b) and Figure \ref{fig:model3_2b} with
$Q_{12} = M_2 =0$ everywhere except for case in Figure \ref{fig:model3_2a}(c) defined by coupling parameters, $c_1 = c_2 = 10$ with $c_1/\ell_1 =100$, where we observe significant distortion throughout the sample. The case in Figure \ref{fig:model3_2a}(c) describes very strong magneto-nematic coupling compared to the other parameters.

It is worth emphasizing that in the ``Laplace limit" with $Q_{12} = M_2 = 0$, the solution profiles for $\Q$ and $\M$ necessarily have points with $\mathbf{Q} = (0,0)$ and $\mathbf{M} = (0,0)$ since $Q_{11}$ and $M_1$ change sign at an interior point $y_Q, y_M \in \left(0, 1 \right)$. For example, if $\mathbf{Q} = \left(-1, 0 \right)$  at $y = 0$, then this describes homeotropic anchoring with $\varphi = \pi/2$ on $y=0$. If $\mathbf{Q} = \left(1, 0 \right)$ at $y=1$, then this describes planar anchoring with $\varphi = 0$ on $y=1$.
Since $Q_{12}$ is identically zero, we must have either $\varphi = 0$ or $\varphi = \pi/2$ everywhere in the domain (refer to Equation~ \ref{eq:16}). At $y_Q \in (0,1)$, we have $Q_{11}\left(y_Q\right) = Q_{12}\left(y_Q\right) = 0 $ so that $\varphi$ is not defined at $y_Q$ and there is a jump discontinuity in the director profile regularised by $|\Q|$$=0$. This jump discontinuity is an example of a domain wall that separates two distinct regions with $\varphi = \pi/2$ on one side of the discontinuity and $\varphi = 0$ on the other side of the discontinuity. In general, $y_Q \neq y_M$ so the domain walls need not be at the same location but numerical results suggest that they occur close to one another. This could be a limitation of the two-dimensional modelling approach since $\mathbf{Q} = (0,0)$ describes a disordered nematic domain wall in three dimensions and one might expect the magnetic nanoparticles to migrate away from the disordered defect wall. Equally, nematic ordering is a long-range effect and whilst the norm of the magnetization vector is small for $|\mathbf{Q}| = 0$, it need not be exactly zero so that some non-zero magnetization is retained. 

\begin{figure}[!htbp]
	\centering
	\includegraphics[scale = 0.15]{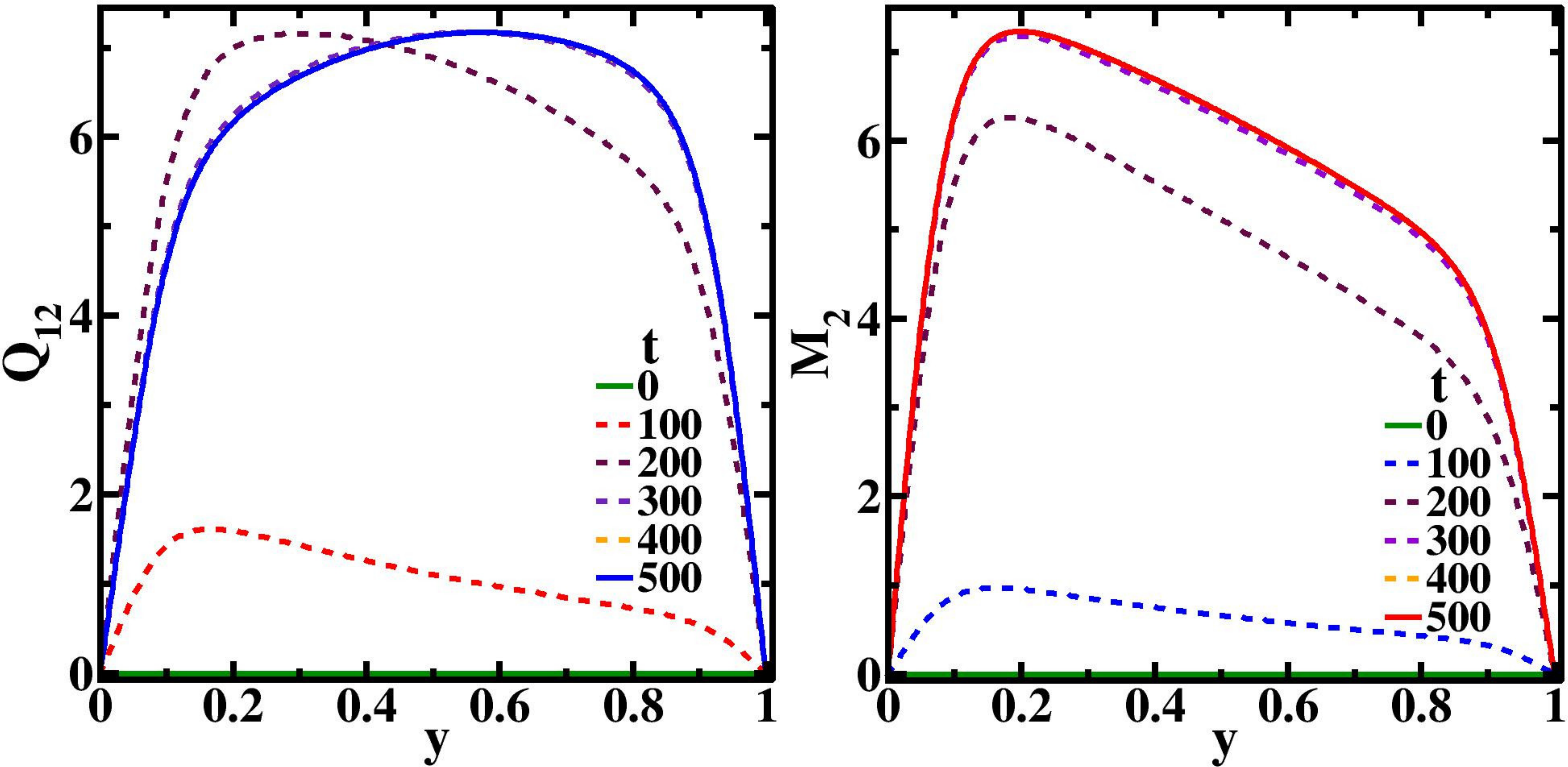}
	\caption{Stability analysis for solutions with non-zero $Q_{12}$ and $M_2$ (in Fig.~ \ref{fig:model3_2a}(c)) by using gradient-flow method defined by (\ref{eq:25}) for parameters $\ell_1 = \ell_2 = 0.1$, $c_1 = c_2 = 10$ and $\mu = 5$. The small perturbation $Q_{12}(y,t=0) = M_2(y,0) = 0.01y(1-y)$ grows and converge to solutions with non-zero $Q_{12}$ and $M_2$ for long-time ($t\sim 500$).}
	\label{fig:model3_4}
\end{figure}
Next, we discuss the stability of the solution branch with $Q_{12} = M_2 = 0$ as a function of the coupling parameters $c_1$ and $c_2$, for the boundary-value problem considered in Figure \ref{fig:model3_2a}. It is clear that we have a solution branch with $Q_{12} = M_2=0$ for all values of $\ell_1, \ell_2, c_1, c_2$. 
The numerics suggest that this solution branch loses stability for large $c_1$ and $c_2$. We can provide some heuristics to this effect. The coupling energy is
$$ - \gamma M_i Q_{ij} M_j = \gamma s |\mathbf{M}|^2 \left( \frac{1}{2} - \cos^2\theta_{\mathbf{n}\mathbf{M}} \right) $$
where $\theta_{\n\M}$ is the angle between the nematic director $\n$ and the magnetization vector $\M$ (also see equations (\ref{eq:15}) and (\ref{eq:16})). The coupling energy is clearly minimized if $\n$ and $\M$ are perfectly aligned with each other. For solutions with $Q_{12} = M_2 = 0$, we have domain walls separating regions with $\varphi =\pi/2$, $\mathbf{M} = \left(-1, 0 \right)$ from regions with $\varphi = 0$, $\mathbf{M} =\left(1, 0 \right)$ and $\mathbf{n}$ and $\mathbf{M}$ are not aligned when $\varphi = \pi/2$ and $\mathbf{M} = (-1, 0)$. As $c_1$ and $c_2$ become larger, the energetic penalty for the mismatch between $\mathbf{n}$ and $\mathbf{M}$ increases and hence, these domain walls are not preferred energetically and we get solutions as in Figure \ref{fig:model3_2a}(c), for which $\n$ and $\M$ tend to align with each other, to minimize the dominant coupling energy. We also observe that in this case, $|\Q|$ and $|\M|$ tend to constants in the middle of the cell, as in the previous model and this constant is greater than the boundary values of $|\Q|$ and $|\M|$. Referring to \cite{luoerbanmajumdar2012}, the maximum principle dictates that the maximum value of $|\Q|$ and $|\mathbf{M}|$ is attained on the boundaries. In the Ginzburg-Landau and Laplace limits of these coupled systems, the system is maximally ordered at the boundaries but for strongly coupled systems as in Figure~\ref{fig:model3_2a}(c), the ordering seems to steadily increase in the bulk. This warrants further investigation and interpretation in the future. 
We illustrate this more conclusively by using the gradient flow model for the free energy in (\ref{eq:17}); the gradient flow model is based on the principle that systems evolve to a state of minimum energy or at least to a local energy minimizer according to the choice of initial conditions \cite{majumdarcanevarispicer2017, gradientflow}. The governing partial differential equations are:
\begin{eqnarray}
\label{eq:25}
&& \mu \frac{\partial Q_{11}}{\partial t} = \ell_1 \frac{d^2 Q_{11}}{dy^2} + Q_{11} - \left(Q_{11}^2 + Q_{12}^2\right) Q_{11} + \frac{c_1}{2}\left(M_1^2 - M_2^2 \right) \nonumber \\ && \mu \frac{\partial Q_{12}}{\partial t} = \ell_1 \frac{d^2 Q_{12}}{dy^2} + Q_{12} - \left(Q_{11}^2 + Q_{12}^2\right) Q_{12} + c_1 M_1 M_2 \nonumber \\ && \mu \frac{\partial M_1}{\partial t} = \ell_2 \frac{d^2 M_1}{dy^2} + M_1 - \left(M_1^2 + M_2^2\right) M_1 + c_2 \left( Q_{11} M_1 + Q_{12} M_2 \right)\nonumber \\
&& \mu \frac{\partial M_2}{\partial t} = \ell_2 \frac{d^2 M_2}{dy^2} + M_2 - \left(M_1^2 + M_2^2\right) M_2 + c_2 \left( Q_{12} M_1 - Q_{11} M_2 \right)
\end{eqnarray}
where $\mu>0$ is a positive constant.
We take $\ell_1 =\ell_2 = 0.1$, $c_1 = c_2 =10$ and $\mu = 5$. This is an initial boundary-value problem; the boundary conditions are as in Figure \ref{fig:model3_2a}. We need to prescribe initial conditions too. For initial conditions with $Q_{12} = M_2 = 0$, the solutions of the system (\ref{eq:25}) have $Q_{12} = M_2 = 0$ for all times. For slightly perturbed initial conditions, for example with $Q_{12}(y, t= 0) = 0.01y(1-y)$ and $M_2 \left(y, t = 0\right) = 0.01y(1-y)$, the solutions distinctly converge to solutions with non-zero $Q_{12}$ and $M_2$ for long-time, as illustrated in Figure \ref{fig:model3_4}.

\begin{figure}[!htbp]
	\centering
	\includegraphics[scale =0.15]{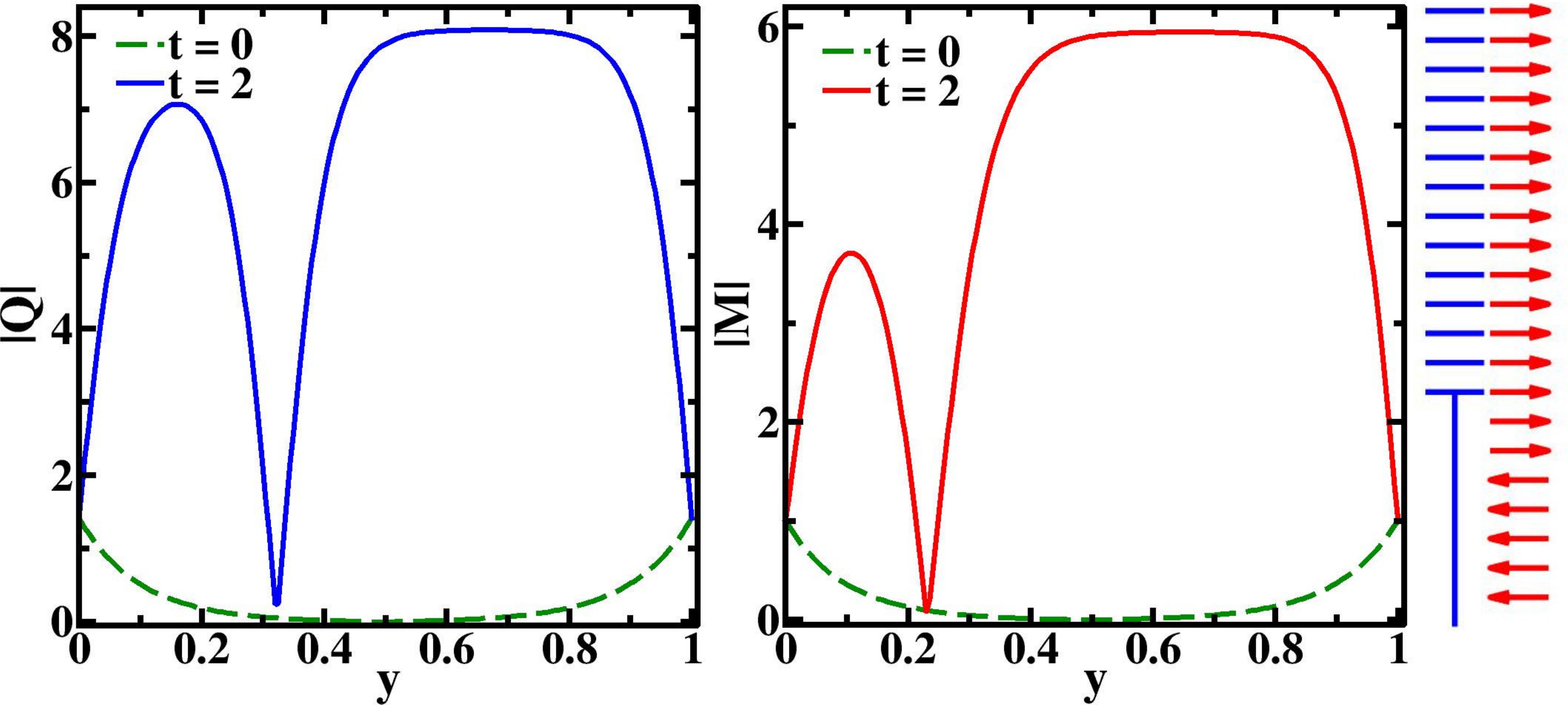}
	\caption{Evolution of $|\Q|$ and $|\M|$ with rapid quenching of the system governed by (\ref{eq:26}) for parameters $\mu =1$, $g=10$, $\ell_1=\ell_2=0.1$ and $c_1=c_2=1$.  The system transform from an initial state of isotropic phase  (small bulk values of $|\Q|$ and $|\M|$) at a temperature $T >$ max\{$T_c^n,T_c^m$\} to a polydomain state for $T <$ min\{$T_c^n, T_c^m$\}. The arrow plots show variation in $\mathbf{n}$ (blue) and $\mathbf{M}$ (red) along the $y$-axis at time $t = 2$.}
	\label{fig:model3_5}
\end{figure}
Finally, we look at the effects of temperature on polydomain formation in such systems. In \cite{Mertelj}, experiments show that fast quenching the sample from an isotropic state in the absence of a magnetic field produces a polydomain sample with two opposing states of magnetization. We try to numerically reproduce the same effect with a variable temperature parameter in the gradient-flow model in (\ref{eq:25}). The temperature parameter is the coefficient of $Q_{11}, Q_{12}, M_1, M_2$ in (\ref{eq:25}); this parameter can be modelled by a function $A(t) = - g\left(2t - 1\right)$ for some positive constant $g$ so that temperature decreases as time increases; lower temperatures favour domain formation with greater ordering i.e. larger values of $|\Q|$ and $|\M|$. $A(t)$ is positive at $t = 0$ so that the initial condition is relatively disordered with small bulk values of $|\Q|$ and $|\M|$ for temperatures $T >$ max\{$T_c^n, T_c^m$\} where $T_c^n$ and $T_c^m$ are critical temperatures for nematic and magnetic materials respectively. As time increases, $A(t)$ becomes negative and this is expected to enhance the degree of ordering and polydomain stabilisation for temperatures $T<$ min\{$T_c^n, T_c^m$\}. For typical materials used in experiments, $T_c^m > T_c^n$ \cite{PH,FK,CK,SS,PC,MH}.
We take the boundary conditions to be $$ Q_{11} = -1, Q_{12} = 0; M_1 = -1, M_2 = 0 \quad \textrm{at $y=0$} $$ and
$$ Q_{11} = 1, Q_{12} = 0; M_1 =1, M_2 = 0 \quad \textrm{at $y=1$.} $$
We work with $\ell_1 = \ell_2 = 0.1$ and $c_1 = c_2 =1$ i.e. relatively small values of the coupling parameter that favour solutions with domain structures (the Laplace limit) as suggested in Figure \ref{fig:model3_5}.
The gradient-flow model with a variable temperature parameter is given by:
\begin{eqnarray}
\label{eq:26}
&&\mu\frac{\partial Q_{11}}{\partial t} = \ell_1 \frac{d^2 Q_{11}}{dy^2} + g\left(2t - 1\right)Q_{11} - \left(Q_{11}^2 + Q_{12}^2\right) Q_{11} + \frac{c_1}{2}\left(M_1^2 - M_2^2 \right) \nonumber\\&&\mu \frac{\partial Q_{12}}{\partial t} = \ell_1 \frac{d^2 Q_{12}}{dy^2} + g\left(2t - 1\right)Q_{12} - \left(Q_{11}^2 + Q_{12}^2\right) Q_{12} + c_1 M_1 M_2 \nonumber \\&&\mu \frac{\partial M_1}{\partial t} = \ell_2 \frac{d^2 M_1}{dy^2} + g\left(2t - 1\right)M_1 - M^2 M_1 + c_2 \left( Q_{11} M_1 + Q_{12} M_2 \right)\nonumber \\
&&\mu\frac{\partial M_2}{\partial t} = \ell_2 \frac{d^2 M_2}{dy^2} + g\left(2t - 1\right)M_2 - M^2 M_2 + c_2 \left( Q_{12} M_1 - Q_{11} M_2 \right)
\end{eqnarray} with $g=10$. The initial condition is a solution of the Euler-Lagrange equations for $A(0)=10$ i.e. a solution of the system
\begin{eqnarray}
\label{eq:27}
&& \ell_1 \frac{d^2 Q_{11}}{dy^2} = 10Q_{11} + \left(Q_{11}^2 + Q_{12}^2\right) Q_{11} - \frac{c_1}{2}\left(M_1^2 - M_2^2 \right) \nonumber \\
&& \ell_1 \frac{d^2 Q_{12}}{dy^2} = 10Q_{12} + \left(Q_{11}^2 + Q_{12}^2\right) Q_{12} - c_1 M_1 M_2 \nonumber \\
&& \ell_2 \frac{d^2 M_1}{dy^2} = 10M_1 + \left(M_1^2 + M_2^2\right) M_1 - c_2 \left( Q_{11} M_1 + Q_{12} M_2 \right) \nonumber \\
&& \ell_2 \frac{d^2 M_2}{dy^2} = 10M_2 + \left(M_1^2 + M_2^2\right) M_2 - c_2 \left( Q_{12} M_1 - Q_{11} M_2 \right)
\end{eqnarray}
subject to the imposed boundary conditions. The numerical solutions for $\Q$ and $\M$ have $Q_{12} = M_2 = 0$ for all times and exhibit distinct points 
$y_{\Q}$ where $\Q$$ = (0,0)$ and $y_{\M} $ where $\M$$=(0,0)$. The point $y_{\M}$ is a domain wall in a three-dimensional sample that separates two distinct domains, with $\M$$ = (-1, 0)$ and $\M$$ = (1,0)$ respectively. For large times, $|\M|$ is larger in magnitude compared to $|\M|$ at $t=0$ i.e. while the polydomain structure is preserved for all times, the degree of ordering or alignment within a polydomain increases with time as temperature decreases, and we observe well-developed polydomains with opposing states of magnetization induced by lowering the temperature or equivalently, the ``quenching" effect.
This relatively simple numerical experiment provides a theoretical and qualitative explanation for the experimentally observed polydomains with opposing magnetizations in \cite{Mertelj}. 

\section{Conclusion}
\label{conclusion}

We study spatial pattern formation in a one-dimensional confined nematic system with suspended magnetic nanoparticles, in three different variational frameworks - (i) the simplest Oseen-Frank framework with constant magnetization; (ii) the Oseen-Frank framework with variable magnetization that allows for domain walls in the magnetization vector and (iii) a two-dimensional Landau-de Gennes framework that allows for domain walls in both the nematic ordering and the magnetization, with Dirichlet conditions for the nematic director and magnetization vector on the boundaries. Two-dimensional Landau-de Gennes models are reduced models that work well for thin geometries; in this case, the three-dimensional domain would be a thin infinite channel whose height is very small compared to the lateral dimensions in the $x$ and $y$-directions, with planar boundary conditions on the bounding surfaces in the $xy$-plane and Dirichlet conditions on the bounding surfaces in the $xz$-plane. This model reduction can be rigorously justified using gamma-convergence techniques \cite{golovaty}. All three models have key re-scaled material-dependent elastic constants and re-scaled magneto-nematic coupling parameters.   We discuss conditions under which the nematic and averaged magnetization profiles follow each other, which could be exploited to pattern nematic configurations using magnetic nanoparticles for desired applications. Models (ii) and (iii) allow for variable order parameters in the magnetization (but not the nematic order parameter) and in both the nematic order parameter and the magnetization respectively; they predict domain walls or phase boundaries defined by $|{\Q}| = |{\M}| = 0$ for coupling constants smaller than or comparable to the elastic constants and in some cases, larger than the elastic constants by an order of magnitude (in the dimensionless framework). The location of these domain walls can be tailored or manipulated by choices of the re-scaled coupling and elastic constants. 

These domain walls qualitatively explain the experimentally observed polydomain structures, for example in \cite{Mertelj}. Using a variable temperature parameter, we also qualitatively explain stable polydomain formation as the temperature is lowered and the polydomain formation is captured in terms of $|\Q|$ and $|\M|$. We numerically verify that these domain walls lose stability as the coupling constants increase in magnitude. It is interesting that domain walls in $\M$ lose stability for $c/l =10$ in the second model (where we assume that $\ell_1 = \ell_2$) whilst they retain stability for $c/\ell=10$ in the third model (assuming that $c_1 = c_2$ and $\ell_1 = \ell_2$). As the domain walls lose stability, we observe more uniform alignment between the nematic director and the magnetization vector and this may correspond to uniformly dispersed suspension of magnetic nanoparticles. The physically relevant range of values for the re-scaled elastic constants and coupling constants ($\ell_1, \ell_2, c_1, c_2$) is clearly heavily dependent on the NLC, the nanoparticles and the temperature. We will explore this in greater detail in future work.

We numerically study pattern formation in a system of magnetic nanoparticles suspended in a nematic medium, in a simplified one-dimensional setting. The models, albeit simplified, illustrate how the alignment between magnetic nanoparticles and the ambient nematic medium can be controlled by the elastic constants and the coupling parameters, to yield different stable textures. Of course, the coupling parameters need to be appropriately interpreted as in \cite{brochard1} but these relatively simple models yield quantitative estimates for the range of values for the re-scaled coupling and elastic constants that allow for polydomain formation. These predictions could be used to interpret future experiments or even design new experiments on suspensions of nanoparticles in nematic media. Equally, we will extend our work to more realistic higher-dimensional models in the future to better capture the physics of the magneto-nematic interactions, both with and without external magnetic fields, with emphasis on manipulating the solution landscapes for desired properties e.g. can nematic defect walls with $|{\Q}|=0$ trap nanoparticles or repel nanoparticles and how can this response be tuned for future applications. We will report on these aspects in future work. 

\section*{Acknowledgments}

KB acknowledges CSIR India for financial support under the grant number 09/086(1208)/2015-EMR-I. A.M. also acknowledges support from an OCIAM Visiting Fellowship and the Keble Advanced Studies Centre. The authors would like to thank the International Centre for Mathematical Sciences where they met for the first time, for follow-on funding and would like to thank DST-UKIERI for generous funding to support this 3-year collaborative project. The authors gratefully acknowledge the HPC facility of IIT Delhi for the computational resources.

\bigskip
\bibliographystyle{elsarticle-num}
\bibliography{ref4}

\end{document}